\documentclass[11pt]{article}

\usepackage[final]{acl}

\usepackage{times}
\usepackage{latexsym}

\usepackage[T1]{fontenc}

\usepackage[utf8]{inputenc}

\usepackage{microtype}

\usepackage{inconsolata}

\usepackage{graphicx}

\usepackage{comment}
\usepackage{amsmath}
\usepackage{algorithm}
\usepackage{algpseudocode}
\usepackage{adjustbox}
\usepackage{booktabs}
\usepackage{multirow}

\usepackage{subcaption}
\usepackage{xcolor}
\usepackage{enumitem}

\usepackage[table]{xcolor}
\usepackage{pifont}
\usepackage{xcolor}

\newcommand{\cmark}{{\color[HTML]{009900}\ding{51}}}
\newcommand{\xmark}{{\color[HTML]{CC0000}\ding{55}}}

\algrenewtext{For}[2]{\algorithmicfor\ #1 \textbf{compute} \ #2}

\title{Dialogue to Discovery: Attribute‑Aware Preference Elicitation for Conversational Product Search Assistants}

\author{
 \textbf{Sarthak Harne\textsuperscript{2$\dagger$}},
 \textbf{Natwar Modani\textsuperscript{1}},
 \textbf{Debabrata Mahapatra\textsuperscript{1}},
 \textbf{Shubham Agarwal\textsuperscript{3$\dagger$}},
\\
 \textsuperscript{1}Adobe Research,
 \textsuperscript{2}Microsoft Research,
 \textsuperscript{3}UC Berkeley,
\\
\textsuperscript{$\dagger$} Work done while at Adobe Research.
\\
 \small{
   \textbf{Correspondence: Natwar Modani} \href{mailto:nmodani@adobe.com}{nmodani@adobe.com}
 }
}

\begin{document}
\maketitle
\begin{abstract}
Conversational product search assistants offer a more expressive, natural, and interactive alternative to traditional keyword-based product search.
With limited screen space, showing only a few items increases the need for precise preference elicitation, which can prolong conversations, leading to user frustration and session abandonment.
Conversely, rushing to recommend items without a clear understanding of preferences risks poor matches and a degraded user experience.
We present Dialogue to Discovery (\textsc{D2D}), an attribute-oriented preference elicitation framework that dynamically exploits the structure of product attributes to efficiently steer conversations toward the user’s desired item. \textsc{D2D} adaptively prioritizes the most informative queries and strategically times product recommendations, reducing premature or off-target suggestions that harm engagement. To evaluate D2D, we curate three datasets from the Amazon Reviews corpus.
In simulated conversations modeled using a multi-factor utilitarian patience framework, D2D achieves a \textbf{22.2–29.9\%} improvement in target-finding accuracy, \textbf{6.6–16.1\%} reduction in abandonment, and \textbf{27.5\%} shorter average conversations
over the state-of-the-art baselines. A complementary user study further confirms significant gains in both user satisfaction and perceived efficiency.
\end{abstract}

\section{Introduction}
\label{sec:intro}
Traditionally, shopping websites~\cite{walmart, amazon, eBay} have relied on keyword-based or semantic search~\cite{robertson2009probabilistic, mearch} to help users find items\footnote{We use the terms `item' and `product' interchangeably} of interest. While reasonably effective,
this approach often returns an large number of results for popular queries. To help users narrow down these results, many platforms~\cite{Faceted} offer facet- or aspect-based filtering. However, this method has its own limitations: users may still be left with an unmanageably large set of items, or conversely, no viable options at all. Moreover, the burden of deciding which filters to apply---and in what order---falls entirely on the user.

With the advent of powerful LLMs~\cite{lin2025can, zhao2024recommender}, \emph{Conversational Product Search (CPS) assistants}~\cite{amazon_rufus, Zou_2022, pscon2025} have become increasingly practical, enabling users to express needs and preferences in a more natural, flexible manner. Recent work has shown that LLMs can serve as effective zero-shot recommenders~\cite{llms-as-zs-recommenders}. However, such systems have their own challenges~\cite{gao2021advances}. One practical issue is that not all user interactions are easier in a conversational setting. As a result, such systems are often deployed in sidebar interfaces~\cite{zendesk_chat, tidio, intercom}, with most of the screen occupied by conventional web content.
This limited real estate makes it impractical to present long result lists,
while extended multi-turn clarifications risk user fatigue and abandonment.
Furthermore, handing over the entire task of finding relevant items for a user query---even after an initial semantic search---to LLMs can incur significant computational cost~\cite{yang2024optimization, shekhar2024towards}, limiting their applicability in the real world.

In this paper, we address the challenge of making conversational search both \textit{efficient}---helping users find desired items quickly---and \textit{engaging}---ensuring dialogue and results remain aligned with user needs and preferences.
The core objective remains the same: successfully guiding the user to their target item.

We present an attribute-based preference elicitation framework, called \textbf{Dialogue to Discovery (\textsc{D2D})}, which dynamically decides whether to recommend a set of products at a given stage of the conversation, and which attributes or values to probe the user about.
\textsc{D2D} leverages the distribution of attribute values within the query-relevant portion of the catalog, as well as the user's known preferences---and the associated uncertainty---over those attributes for this. To decide the recommendations, \textsc{D2D} considers the degree of overlap between scores of the top-ranked items, helping avoid premature or irrelevant suggestions.

Another key aspect of our work is modeling a more \textit{realistic user}. Unlike prior work~\cite{lin2023upsrec, llms-as-zs-recommenders} that assumes users answer all queries indefinitely (subject to a max turns constraint), we explicitly model user \textit{patience}.
While each unsatisfactory turn (e.g., irrelevant question or recommendation) reduces patience, helpful interactions can replenish it. The factors influencing the patience level are inspired by~\cite{jin2021patience}.
A session ends if patience is exhausted, capturing realistic abandonment behavior in online shopping.

To evaluate \textsc{D2D}, we curate $3$ datasets having diverse product categories from the Amazon Reviews dataset~\cite{amazon-reviews}, to emulate realistic buying scenarios. We validate the effectiveness of \textsc{D2D} through both simulations, using our proposed patience-based user model, and human evaluation to assess real-world conversational quality.

Our key contributions are as follows:
(1) We propose \textsc{D2D}, an attribute-aware conversational framework that balances preference elicitation and recommendation well for engaging and efficient product discovery.
(2) We introduce a realistic user simulation model that incorporates the notion of patience that is affected by both conversational quality and suggestion relevance.
(3) We curate $3$ datasets from the Amazon Reviews corpus.
We show that \textsc{D2D} outperforms strong baselines across multiple metrics, and we further validate its effectiveness through human evaluation. We plan to open-source our code and dataset for the broader community.

\section{Problem Formulation}
\label{sec:problem}
We consider the setting of a shopper visiting an online shopping site with a broad idea of the product they are looking for (e.g., a $14$ inch laptop), hereafter called as ``target product", and the specific purchase scenario (e.g., for gaming or productivity). They start by issuing an initial query $c_0$, partially describing the need, which is used to select an initial set of candidate products.
We target scenarios where this initial retrieval returns a large candidate set, making it infeasible to display all results or process them in a single LLM call. We assume the number of distinct attributes is much smaller than the candidate set size—typical in e-commerce settings~\cite{amazon-reviews}.
We assume that the shoppers do not change their preferences in the middle of a session, and respond truthfully when probed about their preference, but may abandon the session if the interaction fails to provide value or maintain engagement (\S~\ref{sec:patience}). We also assume that if recommendations are made in a turn, the user first checks if any of the recommended items match the target product, they accept the recommendation, and the session ends there with success. Else, they proceed to respond to any question asked, subject to the patience model.
The goal of the assistant is to guide the user towards this target item by asking relevant questions to elicit user preferences, and making recommendations, while making sure that the user’s patience does not run out.

In a CPS system, we represent a conversation as a sequence of dialogue turns: the $t^\text{th}$ turn is $(c_t, b_t, r_t)$, where $c_t$ is the customer's utterance, $b_t$ the assistant bot's response, and $r_t$ a (possibly empty) list of recommended items from the catalog $C$.
Each item $i \in C$ is represented as a set of attribute-value pairs $\{(j, v_i(j)) \mid j \in A_i\}$, where $A_i$ is the set of attributes and $v_i(j)$ the value of attribute $j$ for product $i$.
The set of all attributes in the catalog is $A:= \cup_{i \in C} A_i$. We denote $i^* \in C$ as the target item the customer wishes to reach without having full knowledge of it in the beginning.

We denote the user's preference for attribute-value pair $(j,k)$ as $p_{jk}$, the \textbf{\textit{attribute-value preference} (AVP)}, where higher values indicate stronger preference.
We denote uncertainty in AVP as $u_{jk}$ (\textbf{\textit{Attribute Preference Uncertainty}, APU}); and the true user preference lies within $[p_{jk}-u_{jk}, p_{jk}+u_{jk}]$.
The user's preference for item $i$, $p_i$, is a weighted sum of constituent AVPs, with item uncertainty $u_i$ derived from the constituent APUs.

\section{\textsc{D2D} Assistant}
\label{sec:d2d}

Conversational search begins with an initial user query. The assistant first retrieves a catalog subset via embedding similarity; to ensure efficiency, all subsequent processing is limited to this subset.

At each turn $t \geq 1$, the system updates the user Attribute-Value Profile (AVP) using an LLM (\S\ref{subsec:avp}) and updates item preference scores (\S\ref{subsec:itemScore}). Based on these updates, the model makes two decisions: i) to recommend items or not, and if yes, which items; ii) about which attributes/values should a clarifying question be asked (\S\ref{subsec:decision}). These decisions and selected items are passed to the response generation module (\S\ref{subsec:respGen}). This iterative process is summarized in Algorithm~\ref{alg:assistant} in Appendix~\ref{sec:d2dAppendix}, with full prompts available on GitHub\footnote{\url{https://anonymous.4open.science/r/D2D-PromptsAndExamples-248E/}}.

\subsection{Initial Retrieval}
\label{sec:initRet}
When the user issues their first query $c_0$, our system retrieves top-$n$ items from the catalog with the highest cosine similarity with $c_0$ using Sentence-BERT embeddings~\cite{bge-embedding} (item embeddings are precomputed) to form retrieved catalog $\Bar{C}$.
The assistant attempts to elicits user preference over the sets of attributes and their values only within the items in the retrieved catalog $\bar{C}$. We denote this restricted set of attributes as $\Bar{A}:=\cup_{i\in \bar{C}}A_i$.

\subsection{Updating Attribute Value Preferences}
\label{subsec:avp}
We design an LLM-based AVP module $\mathcal{P}$ that updates user preferences for all attribute-value pairs $(j,k)$ at the $t^\text{th}$ turn based on the conversations:
\begin{align}
    [p_{jk}^t, u_{jk}^t] = \mathcal{P}(c_t, D^{t-1}) \quad \forall j \in \Bar{A},\ k \in \Bar{V}_j, \label{eq:avp}
\end{align}
where $p^t_{jk}$ is the preference score,
$u^t_{jk}$ is the uncertainty in the preference score,
$\Bar{V}_j := \{v_j(i)| i \in \Bar{C}\}$ is the set of possible values of attribute $j$ in the retrieved catalog,
$c_t$ is the customer utterance and
${D^{t-1} :=\{(c_l, b_l, r_l)\}_{l=0}^{t-1}}$ is the previous dialog turns. The LLM in $\mathcal{P}$ is prompted with an instruction to provide the preference score as an integer rating, ${p_{ij}^t \in  \{-P, \cdots, P\}}$ for a $P>0$, and the uncertainty as a positive scalar $u_{jk} \in [0,1]$.
The preference values and uncertainties for all attributes and values are allowed to be updated in every turn to account for the evolving preferences of the customer, though,
in practice, only some of these are updated in most of the turns.

\subsection{Item Preference Score Calculation}
\label{subsec:itemScore}
The preference score of item $i$ at the $t^\text{th}$ turn is defined by its query relevance and the AVPs of its constituent elements:
\begin{align}
    p_i^t := \left(1 + \Bar{s}(e_i, e_{c_0})\right) \sum_{j \in A_i,\ k=v_j(i)}\! p_{jk}^t \label{eq:ips}
\end{align}
for all $i \in \Bar{C}$
where $p_i^t$ is the item preference score, $\Bar{s}$ is the similarity score used in retrieval
normalized and scaled over the retrieved catalog  range between $0$ and $\max_{i\in C'} s(e_i, e_{c_0})$. The item score derived from attribute preference part (second term $\sum p_{jk}^t$) is accentuated by the first term $(1+\bar{s})$ for items with higher query relevance.
We utilize the uncertainties in AVP \eqref{eq:avp} in a similar manner to assign uncertainties to the item preference scores of \eqref{eq:ips} as
\begin{equation}
    u_i^t :=  \frac{1 + \Bar{s}(e_i, e_{c_0})}{|A_i|}\!\!\sum_{j \in A_i,\ k=v_j(i)}\!\!\! u_{jk}^t \ \ \ \forall i \in \Bar{C}. \label{eq:ipu}
\end{equation}
The $\bar{s}$ term ensures that if an irrelevant item has similar attributes, they still do not get a high score.

\subsection{Recommendation and Question Decisions}
\label{subsec:decision}
This module makes two critical decisions: 1) whether to recommend any item via $r_t$ or not
; and 2) which attributes/values to ask question about via $b_t$.
We use the confidence intervals of the preferences defined by their scores and uncertainties to make these decisions. Let the item $\tilde{i}$ have the highest item preference score, $\tilde{i} = \mathrm{arg}\max_{i \in \Bar{C}}p_i^t$.
We define \textbf{\textit{top overlapping item} (TOI)} set as items that overlap with the top item to a sufficient degree:
\begin{align}
    \Bar{C}^t := \left\{i\in \Bar{C}\ \middle|\ \frac{\big|\Delta_i \, \cap\, \Delta_{\tilde{i}}\big|}{\big|\Delta_{\tilde{i}}\big|}  \geq \tau \right\}, \label{eq:c_rel}
\end{align}
where $\Delta_i = [p_i^t - \alpha u_i^t,\ p_i^t + \alpha u_i^t]$ is the
confidence interval of $i$'s preference score,
and $\alpha,\tau \in (0,1)$ are hyperparameters.

The assistant decides to recommend items if the top overlapping set size $|\bar{C}^t| \leq m$ (where $m$ represents available recommendation slots); otherwise, it defers recommendation ($r_t = \emptyset$) to seek more information via attribute-based questions.

We allowed the system to potentially ask question to compare items. However, such questions were asked only $6$ times across the $3$ datasets. So, for clarity and brevity, we defer the discussion on workflow for item questions to Appendix~\ref{sec:which}.

To select which attributes to query, we evaluate two factors of informativeness. The first, Attribute Preference Uncertainty (APU), quantifies the ambiguity in the user's preference model (due to AVP). High APU indicates that the assistant is uncertain about the importance a user assigns to an attribute, leading to uncertainty in item preference scores:
\begin{align}
\delta_j = \sum_{i, i' \in C't, i\neq i'} |\Delta{jk} \cap \Delta_{jk'}| \quad \forall j \in \Bar{A}^t \label{eq:apu}
\end{align}
Here, $\Delta_{jk}$ represents the confidence interval ($p_{jk}^t \pm \alpha u_{jk}^t$) for the preference score of a specific attribute-value pair $(j,k)$.The second factor, Attribute Cumulative Entropy (ACE), measures how effectively an attribute’s values differentiate items, particularly among the highest-ranked candidates. We define this using a modified entropy that accounts for item ordering:
\begin{align}
\gamma_j = \sum_{\substack{n = 10, \ \text{step } 10}}^{|\Bar{C}^t|} H\left(f_j^n\right) \label{eq:ace}
\end{align}
where $H(f_j^n)$ is the Shannon entropy of attribute value frequencies within the top-$n$ items. By summing across increasing values of $n$, we prioritize attributes that are most discriminative at the top of the ranking. We rank attributes by
$\delta_j \times \gamma_j$ and select the  top $m$ to guide question generation.

\subsection{Response Generation}
\label{subsec:respGen}

We design an LLM-based Response Generation (RG) module $\mathcal{R}$ that produces a response $b^t = \mathcal{R}(c_t, r_t, q_t)$. This module uses the recommendation list $r_t$ and the
attributes to query from the DM module.The LLM is prompted to elicit user preferences by explaining how specific choices affect product experience. For attribute-based questions, the RG generates either attribute-attribute comparisons (e.g., "screen size vs. RAM") or value-value comparisons (e.g., "HDD vs. SSD"). For item-based questions, it prompts a direct comparison between specific products. Additionally, the module provides a justification to clarify the intent behind each question.

\section{Simulated User Modeling}
\label{sec:userSim}
Given the high cost of human evaluation, we employ a user simulator. Each synthetic user is defined by a profile, a purchase scenario, and a target product. We prompt an LLM to generate profiles (e.g., brand affinity from historical data) and scenarios (e.g., "upgrading storage" for conversational context). To simulate realistically vague initial intent, we use an obfuscated target description: a 3–5 point summary omitting uniquely identifying attributes. The user LLM uses the profile, scenario, and obfuscated description to generate the initial query. In subsequent turns, it accesses the full product description and profile to ensure consistent and grounded responses.

\subsection{Patience Modeling}
\label{sec:patience}
Existing user simulators often engage indefinitely~\cite{wang-etal-2025-muse, lin2023upsrec, llms-as-zs-recommenders} (except for a fixed upper limit), but real users disengage when frustrated. To address this, we introduce patience modeling, an approach underexplored in conversational systems~\cite{jin2021patience, exploring-landscapes, personal-chars}. Our model determines session abandonment based on four factors: \emph{recommendation relevance, attentiveness, cognitive load, and informativeness}~\cite{jin2021patience}. We initialize user patience $\pi^t$ at turn $t=0$ as $\pi^0 = 1$ and update it according to \eqref{eq:patience} after each assistant response $b_t$ based on these four factors, denoted as $\pi_r, \pi_a, \pi_c$ and $\pi_i$ respectively. We omit static UX factors (e.g., persona, usability) as they remain constant across the turns.

\subsubsection{Recommendation Relevance:}
\label{sec:recRel}
This measures how well the recommended items $r_t$ align with the user's target item. We define a relevance function ${\rho: C \times C \rightarrow [0, 1]}$ to compute similarity between any item and the target item $i^*$:
\begin{align}
    \rho(i) = \frac{\sum_{j \in A_i \cap A_{i^*}} \mathbf{1}(v_j(i) = v_j(i^*))}{|A_i \cap A_{i^*}|}, \label{eq:rho}
\end{align}
where $\mathbf{1}$ is the indicator function.
Please note that this is consistent with the modeling assumption that the items' preferences are a weighted sum of attribute value preferences. The recommendation relevance score is defined as $\pi_r = \max_{i \in r_t} \rho(i) - \omega$,
where $\omega \in [0, 1]$ is a threshold hyperparameter, representing the minimum similarity with the target product that is acceptable. For good recommendations, the value of $\pi_r$ would be positive and high, and for poor recommendations, it will be negative.

\subsubsection{Question Inattentiveness:}
\label{sec:attentive}
We measure inattentiveness through two criteria: (1) the assistant should ask about attributes relevant to the target product ($j \in A_{i^*}$), and (2) it should avoid repeating questions already asked. An LLM extracts attributes from the assistant's response, and based on their relevance and redundancy, we assign an inattentiveness score $\pi_a$ using:
\begin{align*}
\pi_a =
\begin{cases}
\beta_{tu}, & \text{relevant and not asked before} \\
\beta_{ta}, & \text{relevant but already asked} \\
\beta_{nu}, & \text{irrelevant and not asked before} \\
\beta_{na}, & \text{irrelevant and already asked}
\end{cases}
\end{align*}
where each $\beta \in [0,1]$ is a hyperparameter. Intuitively, user patience is expected to deplete more when the assistant repeats questions about the same attribute or ones that do not apply to the target item. Therefore, we expect that $\beta_{tu} < \beta_{ta}$;
$\beta_{nu} < \beta_{na}$, and $\beta_{tu} < \beta_{nu}$.

\subsubsection{Complexity and Informativeness:}
This measures how easy the assistant's response is to understand for a typical user and how well it helps the user understand the domain (e.g., product category) and item attributes. To estimate this, we prompt an LLM with the response and ask it to rate complexity and another prompt to rate informativeness, both on a scale of 1 to 10. These scores are then rescaled to $[0, 1]$ and used as the complexity score $\pi_c$ and informativeness score $\pi_i$.

\subsubsection{Overall Patience:}
\label{sec:update}
The above four patience parameters are used to update the use patience as
\begin{align}
    \pi^{t+1} = \pi^{t} + \lambda_r \pi_r - \lambda_a\pi_a - \lambda_c\pi_c + \lambda_i\pi_i \label{eq:patience}
\end{align}
where the $\lambda$'s represent the relative weight of the factors in the overall patience. As high recommendation relevance and high informativeness help enhance user patience, we take them as adding to the patience, but the other two terms having high value deplete the user patience, so we take their contribution as negative.

\section{Experimental Evaluation}

We evaluate \textsc{D2D} for both quality of results and dialogue efficiency across three Amazon Reviews-derived datasets~\cite{amazon-reviews} with different product domains. We first perform experiments using a simulated user setting based on a multi-factor utilitarian patience model.
Second, we conduct controlled user studies to assess human perception of conversation quality and user satisfaction. Together, these experiments provide an extensive validation of our framework.

For our experiments, we consider the setting as described in \S~\ref{sec:problem}. This requires us to have a catalog containing realistic product data, as well as user product purchase data, which is consistent with the product catalog. We need multiple purchases from the same user so that some of the purchases can be used for building the user profile specifying their (initial) preferences, and then subsequent purchases which can be used as the \textit{target} products.

\subsection{Datasets}
\begin{table}[t]
\centering
\caption{Dataset statistics for each product category.}
\label{table:dataset-stats}
\begin{adjustbox}{max width=0.9\textwidth}
\begin{tabular}{lccc}
\toprule
\textbf{Category} & \textbf{\#Items} & \textbf{\#Attributes}\\
\midrule
Electronics & 2009 & 112 \\
Home and Kitchen & 2001 & 143  \\
Sports and Outdoors & 2015 & 104  \\
\bottomrule
\end{tabular}
\end{adjustbox}
\end{table}

To the best of our knowledge, there are no real-world datasets that contain multiple purchases for users with full product details. Typical collaborative filtering datasets only contain anonymized product identifiers, which do not allow for any content-based search.
Therefore, we construct three domain-focused datasets from the Amazon Reviews corpus~\cite{amazon-reviews}.
The dataset has product data
including title, description, and other structured attribute-value pairs parsed from product feature strings. Further, it also has user reviews data, which provides timestamps, star ratings, and review text, which is crucial for simulating a realistic conversation.
While it does not contain product purchase data, we use reviews as a proxy for the purchase of the corresponding product.

\begin{figure}[b]
  \centering
    \includegraphics[width=0.95\linewidth]{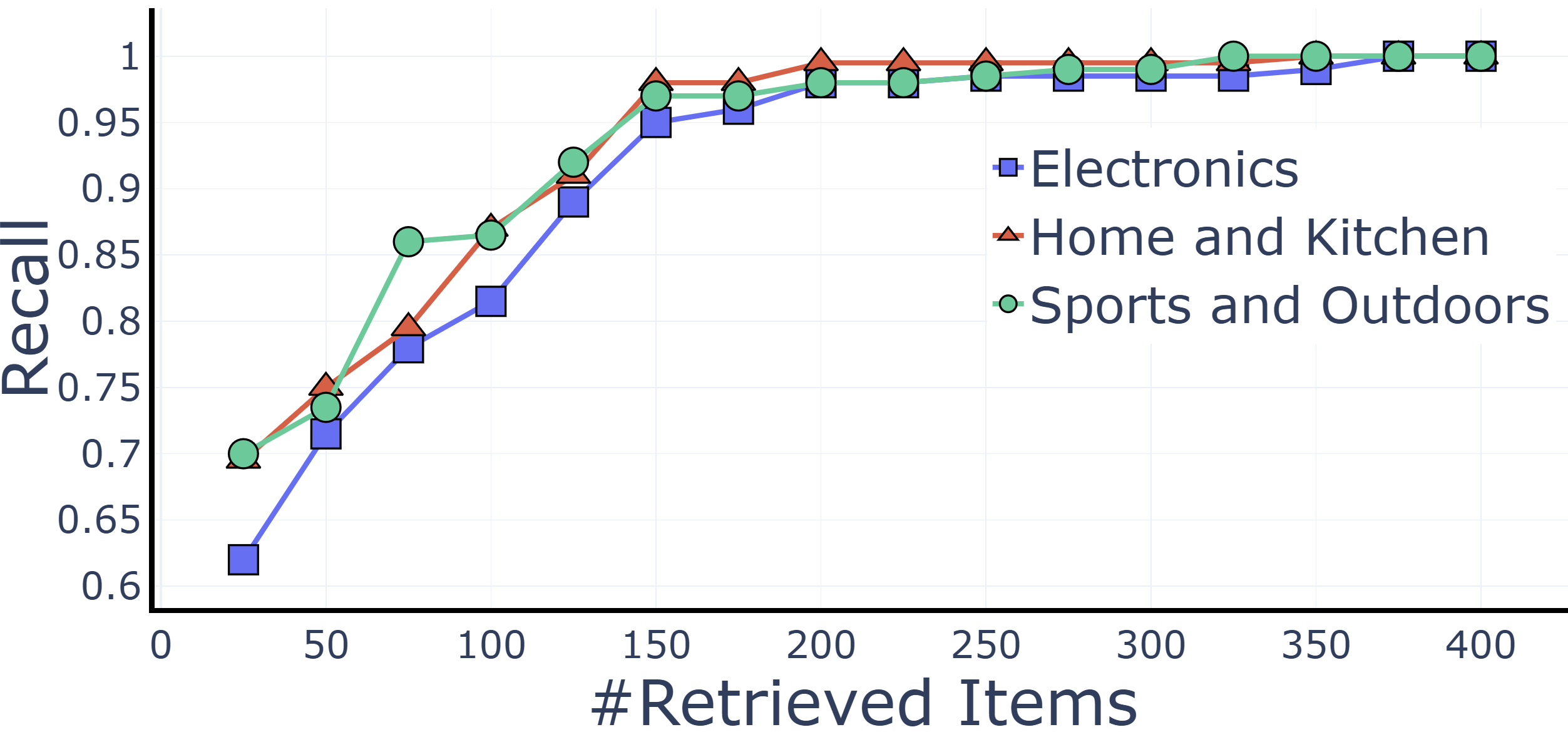}
    \caption{Recall values of the initial retrieval}
    \label{fig:retrieval-recall}
\end{figure}

We selected \textit{Electronics}, \textit{Home \& Kitchen}, and \textit{Sports \& Outdoors} from $28$ categories (Table~\ref{table:dataset-stats}) for their large item counts and product diversity. These categories are also challenging: many products are similar to any given target, requiring a rank cutoff of $200$ to achieve $97\%$ recall in embedding-based retrieval (Figure~\ref{fig:retrieval-recall}).

To construct the dataset for a category, we sort all items by total review count and select around the top $2,000$ items. We then find all users who reviewed at least one of these items. From that group, we select the top $2000$ users based on the number of reviews. Finally, we randomly sample $200$ users from each dataset for experiments.

Each user’s \emph{target item} is set as their most recent positively reviewed item (rating $>3$), while items reviewed before that are used to construct the user profile. For each of these users, we generate the first query/utterance as part of dataset preparation (and to make it standard across methods for fairer comparison).

\subsection{Baselines}
We compare \textsc{D2D} with two type of baselines:

\noindent\textbf{1. Non-conversational retrieval-based methods} simulate traditional keyword search using the first part of the obfuscated target description, e.g., "lightweight water bottle".
We consider two variants: (a) sparse retrieval via BM25~\cite{robertson2009probabilistic}, and (b) dense retrieval using Sentence-BERT embeddings (baai/bge-base-en-v1.5)~\cite{bge-embedding}. Retrieved items are ranked by similarity, and the top-$m$ are directly shown to the user without any dialog.

\noindent\textbf{2. Conversational LLM-based methods} engage in multi-turn interactions with the user. The zero-shot LLM baseline (ZS Recommender~\cite{llms-as-zs-recommenders}) recommends a set of $k$ items at each turn based on dialog history. Since this method does not account for user patience modeling, we extend it by explicitly informing the assistant LLM---via the prompt---that the user may drop off if recommendations or questions are irrelevant or unsatisfactory, instructing it to behave in a patience-aware manner. We call this the \emph{Full-LLM} baseline.

\subsection{Evaluation Metrics}
We evaluate both the quality and efficiency of each method. For quality, we report three metrics: (i) \textbf{Success Rate}, defined as the fraction of sessions where the assistant correctly recommends
the target item; (ii) \textbf{Abandonment Rate}, which captures the fraction of sessions that terminate due to user patience depletion; and (iii) \textbf{NDCG},
two variants, computed based on the recommended items in the last conversation turn. \textit{Binary} variant only considers the target item as relevant and captures the rank of the actual target product. \textit{Fine-grained} variant assign a relevance calculated using a similarity function $\rho$ (Equation~\ref{eq:rho}) between target and recommended items. This captures how similar the recommended items are to the target item in terms of sharing the attribute values.

For efficiency, we report the (i) \textbf{Average Number of Turns} per session, while noting that shorter dialogs may result from either early success or user abandonment. We also measure (ii) \textbf{Token Usage}, tracking both \textit{input (I/P)} and \textit{output (O/P)} tokens per turn and session. Note that input tokens are typically more numerous and less expensive, whereas output tokens are generated by the model, and while fewer, are generally more costly.

In case the system fails to provide parsable output even after two retries, it is considered an error.

\subsection{Settings and Parameter Values}
\label{sec:exp-details}

All methods retrieve top-$200$ items via Sentence-BERT embeddings (baai/bge-base-en-v1.5)
as recall for all three datasets reach very high values ($\geq0.97$) for $200$ retrieved items (see Figure~\ref{fig:retrieval-recall}). All conversations are simulated with \texttt{gpt-4.1-mini} and \texttt{o4-mini}. The simulated \emph{user LLM} modeling the dialogue utterances on behalf of the user remains unchanged and is treated as a \emph{fixed black box} across all techniques. Recommendations per turn are fixed at $m=4$. Scoring uses the overlap threshold $\tau=0.2$ (defined in Eq.~\eqref{eq:c_rel}).
We take $\omega = 0.75$ (\S\ref{sec:recRel}), $\beta_{tu}=0.1,\;\beta_{ta}=0.5,\;\beta_{nu}=0.15,\;\beta_{na}=0.4$ (\S\ref{sec:attentive}) and $\lambda_i=1,\;\lambda_a=1,\;\lambda_c=0.2$, and $\lambda_r=0.05$ (\S\ref{sec:update}) as user-patience modeling hyperparameters.

\subsection{Results: Summary}
\begin{table*}[t]
\centering
\caption{
Evaluation across all datasets (\textbf{DS}) and all methods for two different LLMs. Each row shows results for a retrieval or LLM-based method on a specific dataset for a particular LLM. $\uparrow$/$\downarrow$ indicates whether higher/lower values of the metrics are better.  \textbf{Bold} indicates the best performer in a row; \underline{underline} is for the second-best; `-' indicates that the metric is not applicable.
}
\label{tab:quality-sports}
\begin{adjustbox}{max width=\textwidth}
\begin{tabular}{lll|c|cc|cc|c|cc|cc}
\toprule
 &  &  & \textbf{Success} & \multicolumn{2}{|c|}{\textbf{NDCG $\uparrow$}} & \textbf{Abandonment} & \textbf{Error} & \textbf{Average} & \multicolumn{2}{|c|}{\textbf{Tokens/Turn $\downarrow$}} & \multicolumn{2}{|c}{\textbf{Tokens/Session $\downarrow$}} \\
\textbf{DS} & \textbf{Method} & \textbf{Model} & \textbf{Rate $\uparrow$} & \textbf{Binary} & \textbf{Fine-grained} & \textbf{Rate $\downarrow$} & \textbf{Rate $\downarrow$} & \textbf{\#Turns} & \textbf{O/P} & \textbf{I/P } & \textbf{O/P } & \textbf{I/P} \\
\midrule[0.55pt]

\multirow{8}{*}{\rotatebox[origin=c]{90}{\textbf{Electronics}}}
    & BM25 Retrieval & - & 0.005 & 0.0012 & 0.2376 & - & - & - & - & - & - & - \\
    & Embedding Retrieval & - & 0.025 & 0.0079 & 0.2488 & - & - & - & - & - & - & - \\
    \cline{2-13}

    & ZS Recommender & gpt-4.1-mini & 0.210 & 0.0503 & 0.1284 & 0.835 & 0.065 & 1.67 & 490 & 89167 & \textbf{816} & \textbf{148463} \\
    & ZS Recommender & o4-mini & 0.130 & 0.0319 & 0.0681 & 0.845 & \textbf{0.005} & 2.00 & 674 & 89188 & \underline{1348} & 178398 \\
    & Full-LLM & gpt-4.1-mini & 0.385 & 0.1173 & 0.3165 & 0.560 & 0.035 & 1.80 & 814 & 89464 & 1465 & \underline{161483} \\
    & Full-LLM & o4-mini & 0.240 & 0.0593 & 0.1585 & 0.735 & \textbf{0.005} & 3.04 & 1386 & 90113 & 4180 & 271692 \\
    \cline{2-13}
    & \textsc{D2D} & gpt-4.1-mini & \underline{0.420} & \underline{0.1532} & \underline{0.4198} & \underline{0.520} & 0.040 & 4.96 & \textbf{364} & \textbf{50745} & 1808 & 251903 \\
    & \textsc{D2D} & o4-mini & \textbf{0.500} & \textbf{0.1792} & \textbf{0.5136} & \textbf{0.470} & {\underline{0.010}} & 3.54 & \underline{410} & \underline{51390} & 1456 & 181921 \\

\midrule[0.7pt]

\multirow{8}{*}{\rotatebox[origin=c]{90}{\textbf{Home \& Kitchen}}}

    & BM25 Retrieval & - & 0.050 & 0.0116 & 0.0732 & - & - & - & - & - & - & - \\
    & Embedding Retrieval & - & 0.060 & 0.0130 & 0.0879 & - & - & - & - & - & - & -  \\
    \cline{2-13}

    & ZS Recommender & gpt-4.1-mini & 0.240 & 0.0598 & 0.1102 & 0.690 & 0.065 & 1.43 & 528 & \textbf{83987} & \textbf{755} & \textbf{120101}\\
    & ZS Recommender & o4-mini & 0.125 & 0.0341 & 0.0586 & 0.865 & \textbf{0.005} & 2.13 & 658 & 89194 & 1402 & \underline{189983}\\
    & Full-LLM & gpt-4.1-mini & 0.405 & 0.1072 & 0.2118 & 0.530 & 0.060 & 4.75 & 833 & 85402 & 3962 & 406085 \\
    & Full-LLM & o4-mini & 0.245 & 0.0634 & 0.1047 & 0.740 & \underline{0.010} & 2.56 & 1386 & \underline{84840} & 3540 & 216768 \\
    \cline{2-13}

    & \textsc{D2D} & gpt-4.1-mini & \underline{0.470} & \underline{0.1489} & \underline{0.4090} & \underline{0.515} & \underline{0.010} & 3.54 & \textbf{265} & 91541 & \underline{936} & 323385 \\
    & \textsc{D2D} & o4-mini & \textbf{0.495} & \textbf{0.1667} & \textbf{0.4156} & \textbf{0.495} & \textbf{0.005} & 2.81 & \underline{424} & 92418 & 1192 & 259694\\

\midrule[0.7pt]

\multirow{8}{*}{\rotatebox[origin=c]{90}{\textbf{Sports \& Outdoor}}}

    & BM25 Retrieval & - & 0.050 & 0.0124 & 0.0219 & - & - & - & - & - & - & - \\
    & Embedding Retrieval & - & 0.075 & 0.0173 & 0.0355 & - & - & - & - & - & - & - \\
    \cline{2-13}

    & ZS Recommender & gpt-4.1-mini & 0.355 & 0.1094 & 0.1923 & 0.565 & 0.060 & 1.50 & \underline{391} & \textbf{82576} & \textbf{587} & \textbf{123864} \\
    & ZS Recommender & o4-mini & 0.235 & 0.0751 & 0.1279 & 0.720 & 0.025 & 1.87 & 681 & 89210 & 1274 & \underline{166823} \\
    & Full-LLM & gpt-4.1-mini & \underline{0.460} & 0.1349 & 0.2372 & \underline{0.455} & 0.065 & 3.78 & 476 & 83589 & 1801 & 315966 \\
    & Full-LLM & o4-mini & 0.290 & 0.0791 & 0.1206 & 0.670 & 0.020 & 2.18 & 1249 & \underline{82861} & 2725 & 180636 \\
    \cline{2-13}

    & \textsc{D2D} & gpt-4.1-mini & 0.440 & \underline{0.1359} & \underline{0.3283} & 0.530 & \underline{0.010} & 3.54 & \textbf{276} & 84555 & 978 & 299159 \\
    & \textsc{D2D} & o4-mini & \textbf{0.580} & \textbf{0.2149} & \textbf{0.4796} & \textbf{0.395} & \textbf{0.005} & 2.28 & \underline{391} & 87913 & \underline{891} & 200442 \\

\bottomrule
\end{tabular}
\end{adjustbox}
\end{table*}

We provide a short summary of the results in Table~\ref{tab:quality-sports} here due to space constraints. For a more detailed discussion, please see Appendix~\ref{sec:detailedExpRes}

\textbf{Higher success Rate:} Across all datasets, \textsc{D2D} substantially outperforms the strongest baseline, achieving a mean Success Rate of 52.5\% (vs. 41.7\%) with consistent gains of 22–30\% across domains, and correspondingly large improvements in ranking quality. Mean binary NDCG increases by approximately 56\% and fine-grained NDCG by 84\%, indicating that \textsc{D2D} not only identifies the correct target more often but also ranks highly relevant alternatives near the top. These improvements stem from \textsc{D2D}'s interaction design based on ACE and APU, which focuses on high-impact discriminative attributes preferred by users to effectively disambiguate candidate items.

\textbf{Lower abandonment rate:} \textsc{D2D} lowers the average session abandonment rate from 51.5\% (Full-LLM baseline) to 45.3\% -- a relative reduction of approximately 12\%, with consistent gains across domains. This improvement is driven by better control over when to recommend items and which attributes to probe, avoiding premature recommendations and irrelevant questions.

\textbf{Better token efficiency:} \textsc{D2D} delivers substantial efficiency gains over the Full-LLM baseline, reducing output tokens per session by 51\% and input tokens by 27\% on average for the electronic dataset, and even larger reductions (up to 64\% output and 36\% input) on the other two datasets. These savings stem from AVP-guided, concise queries and overlap-aware recommendation timing that avoids premature or unnecessary item suggestions, reducing both the number and length of generations.

\begin{figure}
    \centering
    \includegraphics[width=0.85\linewidth]{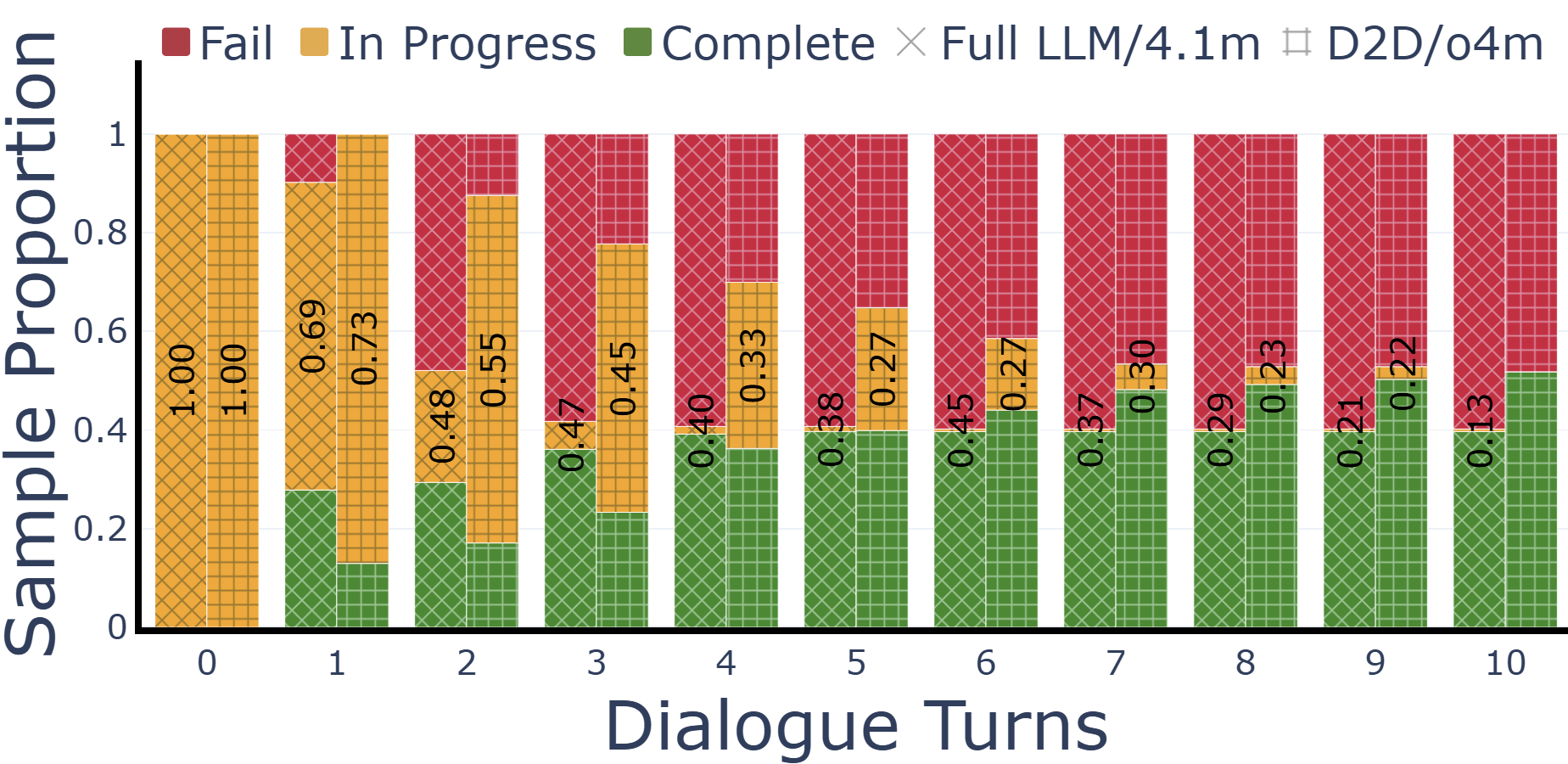}
    \caption{Stacked proportions of conversations ending in success, failure, or still in progress, with average in-progress patience annotation}
    \label{fig:conversation-proportion}
\end{figure}

\textbf{Better user engagement:} Figure~\ref{fig:conversation-proportion} shows that \textsc{D2D} maintains a substantially higher fraction of in-progress conversations across turns and converts them to successes, whereas Full-LLM baselines exhibit early failures, with fewer than 10\% of sessions surviving beyond turn 3. Importantly, \textsc{D2D} preserves higher user patience in in-progress states (e.g., 0.38 vs.\ 0.17 at turn 3), reflecting effective, non-redundant questioning driven by patience modeling and ACE-based attribute selection.

\textbf{Ablations:}
In section~\ref{sec:ablation}, we ablate key \textsc{D2D} components (Table~\ref{tab:ablation-results}), including ACE, APU, TOI, and an LLM-only decision variant (\textsc{D2D-LLM}). Results show that while ACE is most critical, APU and TOI are essential for adaptive questioning and principled recommendation timing, and replacing structured control with LLM decisions consistently degrades performance. This highlights the importance of each component in the overall system.

\subsection{User Study}
\label{sec:userstudy}

We conducted a controlled user study with $50$ subjects (UG and Ph.D. students, and researchers; voluntary participation without payment) to compare chat transcripts from \textsc{D2D} and a Full-LLM baseline for identical queries and target products. Chats were shown side by side in randomized order. Participants answered the following (rephrased for brevity here; full questions and instructions available in Appendix~\ref{sec:userStudyInst}): \textbf{Q1.} Which assistant better elicited your preferences?; \textbf{Q2.} Which assistant timed its recommendations better?; \textbf{Q3.} Which assistant had higher overall conversation quality?

To control for outcome bias, we sampled $50$ conversations spanning four categories based on success outcomes: both \textsc{D2D} and Full-LLM succeed (10), only \textsc{D2D} succeeds (10), only Full-LLM succeeds (10), and both fail (20). These proportions roughly reflect the success distribution observed in our simulation. Participants were unaware of these categories to prevent biasing their responses.

Table~\ref{tab:compact-userstudy} in Appendix reports preferences for each assistant, including ties and abstentions, across Q1-Q3. When one method succeeded and the other failed, the successful method was preferred, as expected. Notably, even when \textsc{D2D} failed and Full-LLM succeeded, users more often favored \textsc{D2D} for overall quality. When both succeeded or failed, \textsc{D2D} was preferred nearly twice as often, reflecting its more targeted questions and better-timed recommendations.

Overall, user preferences consistently favor \textsc{D2D}. These results confirm that \textsc{D2D}’s attribute-aware, patience-driven design not only improves task success in simulation but also leads to higher user satisfaction and perceived quality.

\section{Related Work}
Conversational Product Search (CPS) has progressed from slot-filling dialogue managers~\cite{young2010hidden} and knowledge-based recommenders~\cite{aha2001conversational} to hybrid systems with interactive preference elicitation~\cite{christakopoulou2016towards}. Key challenges remain in orchestrating questions, recommendation timing, dialogue understanding, and evaluation~\cite{gao2021advances}.

Pre-LLM CPS systems focused on attribute selection and dynamic questioning. Attribute-based multi-round strategies~\cite{zhang2018towards}, policy learning under partial observability~\cite{wu2021partially}, and joint modeling of questioning with recommendation triggers~\cite{lei2020estimation} aimed to reduce conversational turns. Knowledge-graph-driven~\cite{ren2021learning} and hierarchical attribute-aware approaches~\cite{wu2023hutcrs} improved multi-attribute disambiguation. While effective in eliciting explicit preferences, none addressed implicit signals or adapted pacing to user patience.

With LLMs, CPS systems shifted toward flexible, open-ended dialogue. Zero-shot recommenders~\cite{llms-as-zs-recommenders} deliver natural conversations but suffer from latency and limited control. Hybrid approaches~\cite{gao2023chat, zheng2024adapting, xu2024llmrecsys} integrate structured user data or collaborative semantics to enhance personalization, while others~\cite{feng2023llmcrs, peng2023implicitllm, lin2023upsrec, li2023intentdriven} focus on sub-task orchestration, implicit preference extraction, or intent-driven recommendations. Retrieval-augmented models~\cite{hou2024bridging} ground outputs in item metadata, improving ranking and factuality. Large-scale multimodal datasets~\cite{wang-etal-2025-muse} and benchmarks~\cite{liang2024llmredial} enable realistic multi-turn evaluation. Yet, explicit mechanisms for coordinating information-seeking questions with recommendations in a user-patience–aware manner remain scarce.

Evaluation has evolved from offline metrics like NDCG~\cite{jarvelin2002cumulated} to LLM-based user simulations~\cite{wang2023rethinking, chen2025recusersim} and human studies on pacing~\cite{jin2021patience, kostric2025should}, showing that poorly timed probing can drive abandonment~\cite{Diriye2012}. Jointly optimizing attribute prioritization, recommendation timing, and user-patience modeling is thus critical~\cite{Zou_2022}.

We address this gap by combining AVP-based preference modeling, APU-guided attribute prioritization, and TOI-aware recommendation timing. Our framework, \textsc{D2D}, improves ranking, reduces abandonment, and maintains efficiency via short, targeted LLM calls, achieving higher success rates across domains.

\section{Conclusion}
\label{sec:conclusion}
We introduced \textbf{Dialogue to Discovery (\textsc{D2D})}, an attribute-aware CPS framework integrating fine-grained preference elicitation with informed recommendation timing. By combining AVP-based uncertainty modeling, ACE-guided attribute selection, and overlap-aware recommendation pacing, \textsc{D2D} ensures that each turn targets the most informative signals while avoiding premature recommendations. Across three Amazon-derived domains, \textsc{D2D} yields shorter conversations, lower abandonment rates, and
higher target-finding success over
several strong baselines. A user study further confirms that human users prefer \textsc{D2D} for its precise queries, well-timed recommendations, and overall dialogue quality.
This demonstrate that structured, attribute-driven, and patience-aware dialogue design can outperform monolithic LLM approaches in both effectiveness and efficiency.

While \textsc{D2D} excels in precision and efficiency, it relies on static patience and attribute models, limiting responsiveness to evolving user behavior and dialogue dynamics.
\emph{Future work} could enhance \textsc{D2D} by: (i) extending patience modeling through user-adaptive decay policies using reinforcement learning from behavioral signals; (ii) enriching AVP scoring with multimodal attribute semantics; and (iii) scaling ACE via hierarchical partitioning of catalogs using attribute taxonomies.

\newpage
\section*{Limitations}
\textbf{Limitations of Evaluation.} Simulated user studies may not capture the full diversity of real-world behaviors, including cultural and accessibility considerations. Human evaluation involved a limited participant pool and may not generalize across demographics. We report aggregate metrics and avoid subgroup-specific claims.

\textbf{Choice of Language and LLM.} Our results are limited to English language and GPT family of LLMs.

\textbf{Reproducibility:} \textit{We provide all (attribute selection, response generation, and simulated user queries) prompts in our experiments and input/output examples in \url{https://anonymous.4open.science/r/D2D-PromptsAndExamples-248E/}.}

\section*{Ethical Considerations}

This work develops Dialogue to Discovery (\textsc{D2D}), an attribute-aware conversational product search framework evaluated in simulated and human settings. We address potential ethical aspects, societal impact, and mitigation strategies.

\textbf{Privacy and Data Usage.} All datasets are derived from publicly available portions of the Amazon Reviews corpus~\cite{amazon-reviews}, using only product metadata (titles, descriptions, structured attributes) and user review texts to construct synthetic profiles and conversations. No personally identifiable information (PII) beyond public usernames is used, and no de-anonymization was attempted. User simulations are generated via large language models (LLMs) without incorporating private or confidential data. Our use of this data is consistent with the MIT License under which this dataset was made available.

\textbf{Fairness and Bias.} The framework relies on structured product attributes and preference signals from LLM outputs, which may reflect biases in the Amazon Reviews dataset (e.g., overrepresentation of certain brands or categories) or in pretrained LLMs. Such biases could affect attribute selection or item exposure. Our evaluation emphasizes accuracy and efficiency, but fairness-aware modeling remains an important direction for future work.

\textbf{Potential Misuse.} D2D is intended to enhance efficiency and engagement in legitimate shopping contexts. Similar techniques could be adapted for manipulative recommendations, potentially influencing user decisions without informed consent. Mitigation requires careful application scoping, transparency in recommendation rationale, and user controls for personalization.

\textbf{Mitigation Strategies.} Released data and code are anonymized and restricted to non-commercial research. We encourage incorporating fairness audits, privacy safeguards, and explainability mechanisms when deploying similar systems. To ensure reproducibility, we provide an anonymized repository \footnote{\url{https://anonymous.4open.science/r/D2D-PromptsAndExamples-248E/}} with prompts and input/output samples; other components are placeholdered for anonymity and will be released in full upon acceptance.

\bibliography{bibliography}

\newpage

\appendix

\section*{Appendix}
\label{sec:appendix}
First, we provide some additional information about the method discussed in Section~\ref{sec:d2d}. We first discuss the item type questions as briefly mentioned in Section~\ref{subsec:decision}. We will then provide an algorithmic picture for the D2D Assistant (Section~\ref{sec:d2d}) as Algorithm~\ref{alg:assistant}, followed by an algorithmic picture of user simulator (Section~\ref{sec:userSim}) as Algorithm~\ref{alg:us}. We then discuss additional details and observations about our experimental results.

\section{D2D Assistant}
\label{sec:d2dAppendix}

\subsection{Item Question}
\label{sec:which}
We allow D2D Assistant to ask the user a question which requires them to compare two items. Here is how we decide if the D2D Assistant should ask an item question.

If we decide not to recommend any items, and the number of top overlapping items in $\Bar{C}^t$ \eqref{eq:c_rel} is no more than $2 m$, and there exists a pair of items that do not have too many different attributes (denoted by hyperparameter $m'$), then we ask an item comparison type question from the user.
Formally, if the set of candidate item-pairs
\begin{align}
    S := \{(i, i')\in \Bar{C}^t \ | \; |A_i \oplus A_{i'}| \leq m' \} \label{eq:cip},
\end{align}
where $\oplus$ is the difference between the union and intersection of the two sets, is non-empty, i.e., $|S| > 0$, then the assistant decides to ask a question on the item pair in $S$ that has the highest preference overlap $\max_{(i, i') \in S} |\Delta_i \cap \Delta_{i'}|$.

The rationale for this decision making structure is that if the top $m$ items have clear score differences from others, the likelihood of a user preferring one of them is high, so we proceed with the recommendation. A set of recommendations are implicitly asking the user if they prefer any of the items shown as recommendation, and therefore, asking another item question is not appropriate.

When we decide to not recommend any items, an item-based question can reveal information about multiple attributes at once; therefore, we look for opportunities to ask an item question. However, if we ask an item question when uncertainty about user preferences is high, we may ask about irrelevant items and fail to learn the user's preference for the actual target item. Thus, we do not ask item questions if the top overlapping item set exceeds $2 m$, and ask attribute questions instead.

\begin{algorithm}[t]
\caption{D2D Assistant in $t^\text{th}$ turn} \label{alg:assistant}
\begin{algorithmic}[1]
\Statex \hspace*{-\algorithmicindent} \textbf{Input:}
Customer utterance $c_t$, past dialogues $D^{t-1}$,
\Statex retrieved catalog $\bar{C}$
\Statex \hspace*{-\algorithmicindent} \textbf{Output:} Bot's response $b^t$, recommendation $r_t$

\State Obtain all attributes $\Bar{A}$ of the retrieved catalog $\bar{C}$

\State \textbf{for each} attribute $j \in \Bar{A}$ and value $k \in \bar{V}_j$ \textbf{compute}
\State \quad \textbf{attribute-value preferences} (AVP):
\State \quad \quad scores and uncertainties $p_{jk}^t, u^t_{jk}$ from \eqref{eq:avp}

\State \textbf{for each} item $i$ \textbf{compute}
\State \quad \textbf{item preferences}:
\State \quad \quad scores and uncertainties $p_i^t, u_i^t$ from \eqref{eq:ips} and \eqref{eq:ipu}

\State Obtain \textbf{top overlapping items} (TOI): $\Bar{C}^t$ from \eqref{eq:c_rel}
\State Obtain \textbf{candidate item pairs}: $S$ from \eqref{eq:cip}

\State \textbf{if} ($|S| > 0$) and $(m < |\Bar{C}^t| < 2m)$ \Comment{item based question}
\State \quad $q^t \gets $ top two overlapping items from $S$
\State \textbf{else} \Comment{attribute based question}
\State \quad \textbf{if} $|\bar{C}_t| \leq m$
\State \quad \quad Recommended items $r_t \gets \bar{C}_t$
\State \quad \textbf{else}
\State \quad \quad $r_t \gets \phi$
\State \quad \textbf{for all} attribute $j\in \Bar{A}$ \textbf{compute}
\State \quad \quad \textbf{attribute preference uncertainties} (APU): $\delta_j$ in  \eqref{eq:apu}
\State \quad \quad \textbf{attribute cumulative entropy} (ACE): $\gamma_j$ in \eqref{eq:ace}
\State \quad \quad \textbf{attribute informativeness}: $\delta_j * \gamma_j$
\State \quad $q^t \gets$ $m$ most informative attributes

\State generate response $b_t \gets R(c_t, r_t, q_t)$

\end{algorithmic}
\end{algorithm}

\begin{algorithm}[h]
\caption{User Simulator in $t^\text{th}$ turn}
\label{alg:us}
\begin{algorithmic}[1]
\Statex \hspace*{-\algorithmicindent}\textbf{Input:} bot utterance $b_t$, recommended items $r_t$, dialog history $D^t$, target item $i^*$
\Statex \hspace*{-\algorithmicindent}\textbf{Output:} customer utterance $c_{t+1}$, status
\State \textbf{if} $i^* \in r_t$ \textbf{then}
\State \quad status $=$ Success
\State \textbf{else if} $t>\mathrm{maxTurns}$ \textbf{then}
\State \quad status $=$ Failure
\State \textbf{else} \Comment{update patience}
\State \quad Compute new patience from \eqref{eq:patience}
\State \quad \textbf{if} new patience $> 0$ \textbf{then}
\State \quad \quad Generate response $c^{t+1}$
\State \quad \textbf{else}
\State \quad \quad status $=$ Failure
\end{algorithmic}
\end{algorithm}

\section{Experimental Results}
\label{sec:detailedExpRes}

\subsection{Results: Quality}
\label{sec:performance}
We compare \textsc{D2D} (with two different LLMs) against both non‑conversational (BM25, Dense Embedding) and conversational (\emph{ZS Recommender}, \emph{Full-LLM}, each with two different LLMs) baselines across the three datasets in Table~\ref{tab:quality-sports}. For all LLM-based methods (including ours), responses are generated under identical prompting logic as far as possible.

We see that the results across the three datasets largely have very similar trends. Another point worth noting is that while for the \textsc{D2D} family of methods, using o4-mini LLM gives the best results, for the LLM-based baselines, using gpt-4.1-mini provides better performance. We believe that this is due to the length of each input that we use in the LLM-based baselines is much closer to the maximum context length of o4-mini (128,000 tokens) as compared to gpt-4.1-mini (1,000,000 tokens), leading to worse performance while using o4-mini as supported by~\cite{leave-no-doc-behind}. On the other hand, each LLM call in D2D uses much less of the LLM's maximum context length (the \#tokens per turn in Table~\ref{tab:quality-sports} for \textsc{D2D} are for 2 LLMs put together, but for the LLM baselines, it is for a single call), leading to better performance in o4-mini.

\paragraph{Reduced Session Abandonment}
Table~\ref{tab:quality-sports} shows that \textsc{D2D} significantly reduces average session abandonment rate from $51.5\%$ (Full-LLM best baseline) to $45.33\%$ ($\approx12\%$ reduction), with domain-specific reductions of $16\%$ in \textit{Electronics}, $13.2\%$ in \textit{Sports \& Outdoors}, and $6.6\%$ in \textit{Home \& Kitchen}. Users abandon sessions due to poor recommendations or being probed repeatedly about the same and/or irrelevant attributes. While we do instruct the `Full-LLM' baselines not to recommend prematurely or ask about (irrelevant) attributes repeatedly, they still end up making these mistakes.
However, \textsc{D2D} allows a much better control over the decision-making regarding when to recommend, resulting in much fewer session abandonment caused by poor recommendations. Also, separating the process of finding attributes, about which the user should be probed, allows easier control over not having repeated questions on attributes.
The overlap-aware recommendation strategy defers item suggestions until confidence intervals converge sufficiently, preventing irrelevant or premature item recommendations (see~\S\ref{subsec:decision}).

\paragraph{Higher Success Rate \& NDCG}
Table~\ref{tab:quality-sports} shows that \textsc{D2D} achieves a mean Success Rate across datasets of $52.5\%$, about $25\%$ higher than that of the strongest baseline (Full‑LLM at 41.67\%). Gains are consistent across domains: about $29.9\%$ in \textit{Electronics} (from $38.5\%$ to $50.0\%$), $~22\%$ in \textit{Home \& Kitchen} (from $40.5\%$ to $49.5\%$), and $~26\%$ in \textit{Sports \& Outdoors} (from $46.0\%$ to $58.0\%$).
These gains in target accuracy are also mirrored in the ranking metrics, as shown in the NDCG column of Table~\ref{table:dataset-stats}. Across datasets, mean binary NDCG improves by $\approx56\%$ (0.1198→0.1869), and fine-grained NDCG by $\approx84\%$ (0.2552→0.4696), indicating that in terms of ranking the products, \textsc{D2D} performs even more impressively. Further, even when the exact target is not top-ranked, \textsc{D2D} reliably surfaces highly similar alternatives. This is crucial in real-world shopping scenarios, where high-relevance alternative items can still satisfy user intent, as they may not have a specific item in mind.

Importantly, these gains arise not just from better ranking, but from better interaction design, reflected in two key design choices in \textsc{D2D}: (i) ACE-bases prioritization of attributes to probe the user using the discriminatory power of the attribute, particularly towards the top of the item ranking, and (ii) APU-based preference modeling, which enables the assistant to distinguish between attributes cognitively salient to the user and having high uncertainty. This allows early focus on high-impact attributes that disambiguate between candidate items.

\subsection{Results: Efficiency}
In addition to accuracy and engagement gains, \textsc{D2D} offers substantial efficiency improvements in LLM usage, as shown in the last two columns of Table~\ref{tab:quality-sports}. Compared to `Full-LLM', output tokens per session averaged over all categories drop by 51\% (2409→1180), while input tokens drop by 27.33\% (294,511→214,019). This is despite higher token usage for the electronic dataset, where the `Full-LLM' baseline makes overconfident recommendations, which leads to a significant session abandonment (and in some cases, also early successful recommendations), resulting in very short sessions ($1.8$ turns per session). For the other two datasets, the reduction in tokens per session is 64\% for O/P tokens and 36.3\% for I/P tokens. Please note that here, the baseline and best proposed methods taken for comparison are the ones achieving the highest success rate, and not the ones with the lowest token usage.
The token per turn are generally comparable across \textsc{D2D} and the conversational baseline, however, please note that the baselines make one LLM call per turn, whereas \textsc{D2D} makes one call for determining attribute value preferences and one for response generation.

These gains arise from three design choices: first, AVP-guided queries are short and semantically focused, avoiding verbose or generic prompts; second, overlap-aware recommendation timing defers costly item suggestions until the candidate space is well-separated; finally, delaying item suggestions until confidence intervals are tight prevents premature recommendations. Together, these mechanisms reduce the number and size of assistant generations without compromising dialog quality. For practical deployment, where LLM inference cost and latency are key constraints, \textsc{D2D} offers higher recommendation effectiveness and without compromising system efficiency.

\textsc{D2D} exemplifies how structured dialog planning, grounded in user modeling and catalog semantics, can simultaneously improve recommendation quality, reduce abandonment, and lower inference cost. The key takeaway is: principled interaction design, not merely larger models, is key to building effective LLM-based assistants.

\subsection{User Patience Analysis}
\label{sec:patienceAnalysis}

\begin{figure}[b]
    \centering
    \includegraphics[width=0.85\linewidth]{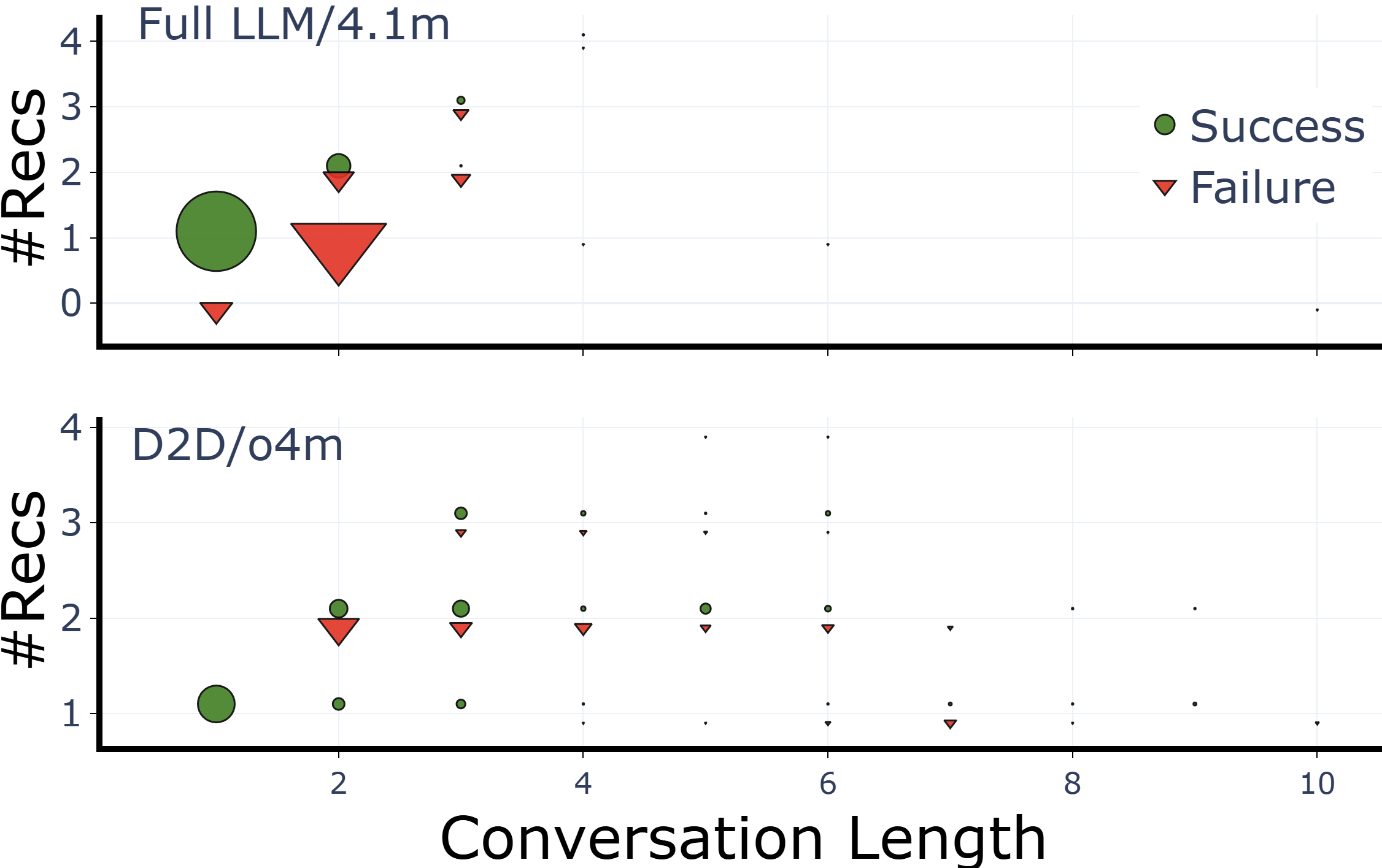}
    \caption{Conversations with their total dialogue length and number of recommendations in that session, size of the markers show the number of sessions with such characteristics}
    \label{fig:n-recs-conv-length}
\end{figure}

To better understand how \textsc{D2D} sustains user engagement, we analyze the evolution of user patience across turns and compare conversation trajectories across success, failure, and in-progress outcomes.

\paragraph{Sustained Engagement Across Turns.}
Figure~\ref{fig:conversation-proportion} tracks the distribution of conversation states: success, failure, or in-progress across turns, with in-progress segments annotated with average patience levels.
Full‑LLM methods exhibit sharp rises in failure rates as early as turn 2, with less than 10\% of conversations surviving till turn 3. In contrast, \textsc{D2D} sustains a significantly higher proportion of in-progress conversations through mid-turns and consistently converts them into successes.
Notably, \textsc{D2D} maintains higher patience even in in-progress states (e.g., 0.38 vs 0.17 at turn 3), highlighting the assistant’s ability to maintain user interest through targeted, non-redundant queries.
This again supports the claim that explicitly informing the assistant—via its prompt—about the user’s patience model, combined with product ranking-aware attribute selection (via ACE) for asking questions to the user, and principled recommendation timing, collectively preserve engagement while steering towards the desired item.

\paragraph{Better Trajectory of Conversation Outcomes.}
Figure~\ref{fig:n-recs-conv-length}
shows a fine-grained picture of conversation trajectories. It simultaneously shows the conversation length (x-axis), number of times recommendations shown in the session (y-axis), the outcome (color and shape of marker) and how many sessions exhibited these characteristics (size of the marker). Note that large number of sessions have recommendations in large fraction of turns for `Full LLM', and the recommendations are highly hit-or-miss (notice large red triangles for conversation length 2). In contrast, \textsc{D2D} tries to understand the user needs before making recommendation in most cases, and therefore, no large red triangles in lower part of the plot, reinforcing the principled timing of recommendation as a key reason for higher success and lower abandonment.

\subsection{Component-wise Ablation}
\label{sec:ablation}

\begin{table*}[t]
\centering
\caption{Component-wise ablation of \textsc{D2D}.
The complete \textsc{D2D} system (top row) performs {best overall}; removing any component reduces success rate, ranking quality, or efficiency. The most damaging ablation (removing ACE) is shaded in {red}. Here, ``(w/o)'' denotes ``without'', ↑ indicates higher is better, and ↓ indicates a lower metric value is better.}
\label{tab:ablation-results}
\begin{adjustbox}{max width=0.99\textwidth}
\begin{tabular}{l|c|cc|c|c|c|cc|rr}
\toprule
 & \textbf{Success} & \multicolumn{2}{|c|}{\textbf{NDCG $\uparrow$}} & \textbf{Abandonment} & {\textbf{Error}}  & \textbf{Average} & \multicolumn{2}{|c|}{\textbf{Tokens/Turn $\downarrow$}} & \multicolumn{2}{|c}{\textbf{Tokens/Session $\downarrow$}} \\
\textbf{Method} & \textbf{Rate $\uparrow$} & \textbf{Binary} & \textbf{Fine-grained} & \textbf{Rate $\downarrow$} & \textbf{Rate $\downarrow$} & \textbf{\#Turns} & \textbf{O/P} & \textbf{I/P} & \textbf{O/P} & \textbf{I/P} \\
\midrule
\textsc{D2D} & 0.500 & 0.1792 & 0.5136 & 0.470 & 0.010 & 3.54 & 410 & 51390 & 1456 & 181921 \\
\hline
\rowcolor{red!10}
(w/o) ACE & 0.390 & 0.1227 & 0.3781 & 0.585 & 0.005 & 3.74 & 415 & 51512 & 1552 & 192655 \\
(w/o) APU & 0.470 & 0.1612 & 0.4672 & 0.500 & 0.010 & 4.06 & 410 & 51410 & 1665 & 208726 \\
(w/o) TOI & 0.415 & 0.1483 & 0.4183 & 0.565 & 0.000 & 5.10 & 407 & 51481 & 2075 & 262553 \\
\textsc{D2D-LLM} & 0.445 & 0.1414 & 0.3901 & 0.515 & 0.020 & 2.33 & 556 & 55970 & 1240 & {129850} \\
\bottomrule
\end{tabular}
\end{adjustbox}
\end{table*}

We ablate each core component of \textsc{D2D} in Table~\ref{tab:ablation-results}. In particular, for attribute-based question selection, we ablate with respect to APU and ACE to form the final attribute informativeness score. For recommendation timing, we ablate the TOI-based condition.
One may ask if an LLM will make good decisions about which items to recommend and what question to ask on each turn, given the AVP framework. Hence, as an additional ablation, we ask an LLM to make both these decisions, where the input is updated attribute value preference scores and item utility scores. We call this approach as \textsc{D2D-LLM}.

\textbf{ACE Ablation.}
Removing ACE means not having knowledge of discriminatory power of the attributes, and results in the largest performance drop: Success Rate falls from 50.0\% to 39.0\%, binary NDCG from 0.1792 to 0.1227, and abandonment rises to 58.5\%. Without ACE, the assistant asks less informative or redundant questions, leading to increased user drop-off due to reduced patience.

\textbf{APU Ablation.}
Without APU, planning becomes less adaptive, relying on static heuristics. This reduces the Success Rate to 47.0\% and the binary NDCG to 0.1612. Turns and token usage also increase, reflecting less efficient elicitation.

\textbf{TOI Ablation.}
Here, we replace the overlap-aware recommendation logic by randomly deciding to recommend products in a turn with a probability equal to the expected number of recommendations during regular operation (p=0.26).
Disabling overlap-aware recommendation causes premature, low-confidence suggestions, lowering the Success Rate to 41.5\% and increasing abandonment to 56.5\%. Fine-grained NDCG remains relatively strong (0.4183), but dialog length (5.1 turns) and token usage (262K per session) rise.

\textbf{D2D-LLM.}
While providing the same framework, but asking an LLM to make decisions (instead of \S~\ref{subsec:decision}) leads to around 11\% reduction in success rate(0.5 to 0.445) and corresponding changes in NDCG. The abandonment rate also rises by 9.5\%. This shows that numeric value based reasoning of the LLM is not able to match performance of the fine-grained control that our method provides.

\subsection{User Study: Instructions and Questions}
\begin{figure*}[!h]
    \centering
    \includegraphics[width=0.85\linewidth]{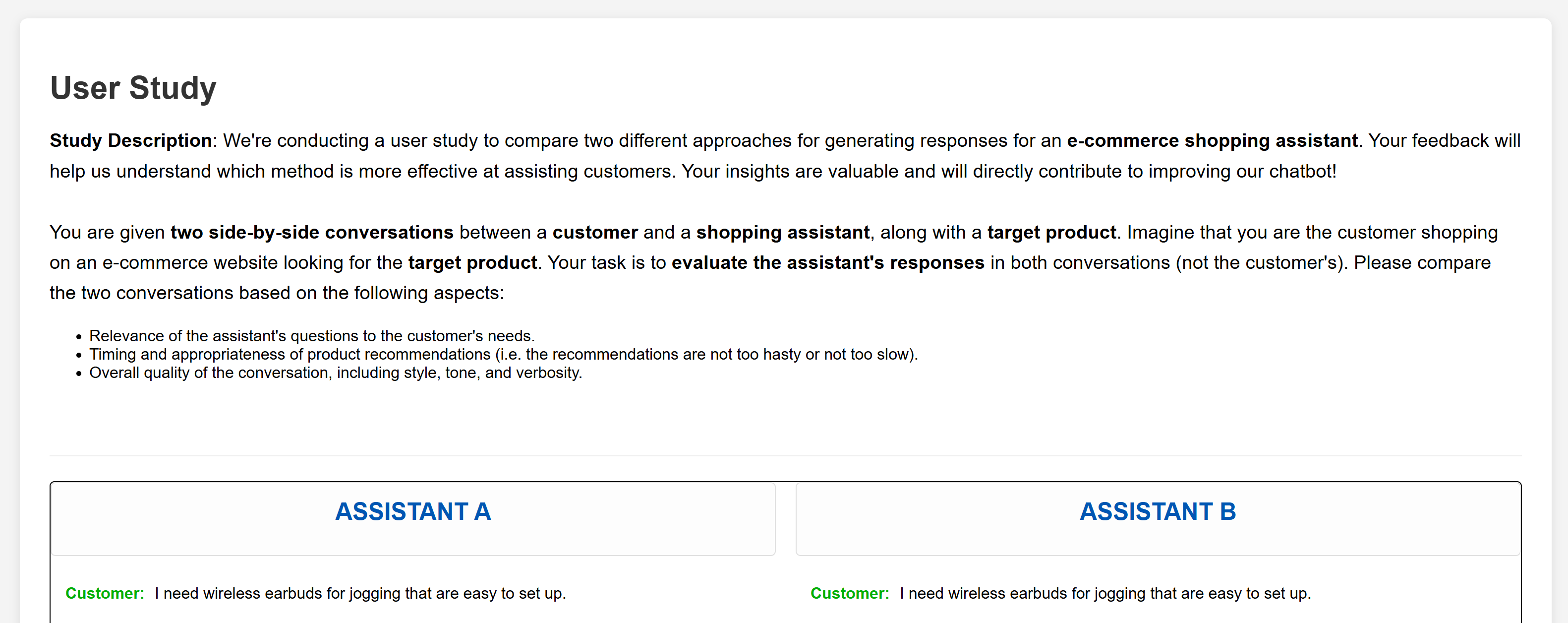}
    \caption{The participants are informed about the intent of the study and explained what they need to do. The assistants are randomized and the users are not aware of the inner mechanisms of either assistant.}
    \label{fig:userstudy-description}
\end{figure*}

\begin{figure*}[!h]
    \centering
    \includegraphics[width=0.85\linewidth]{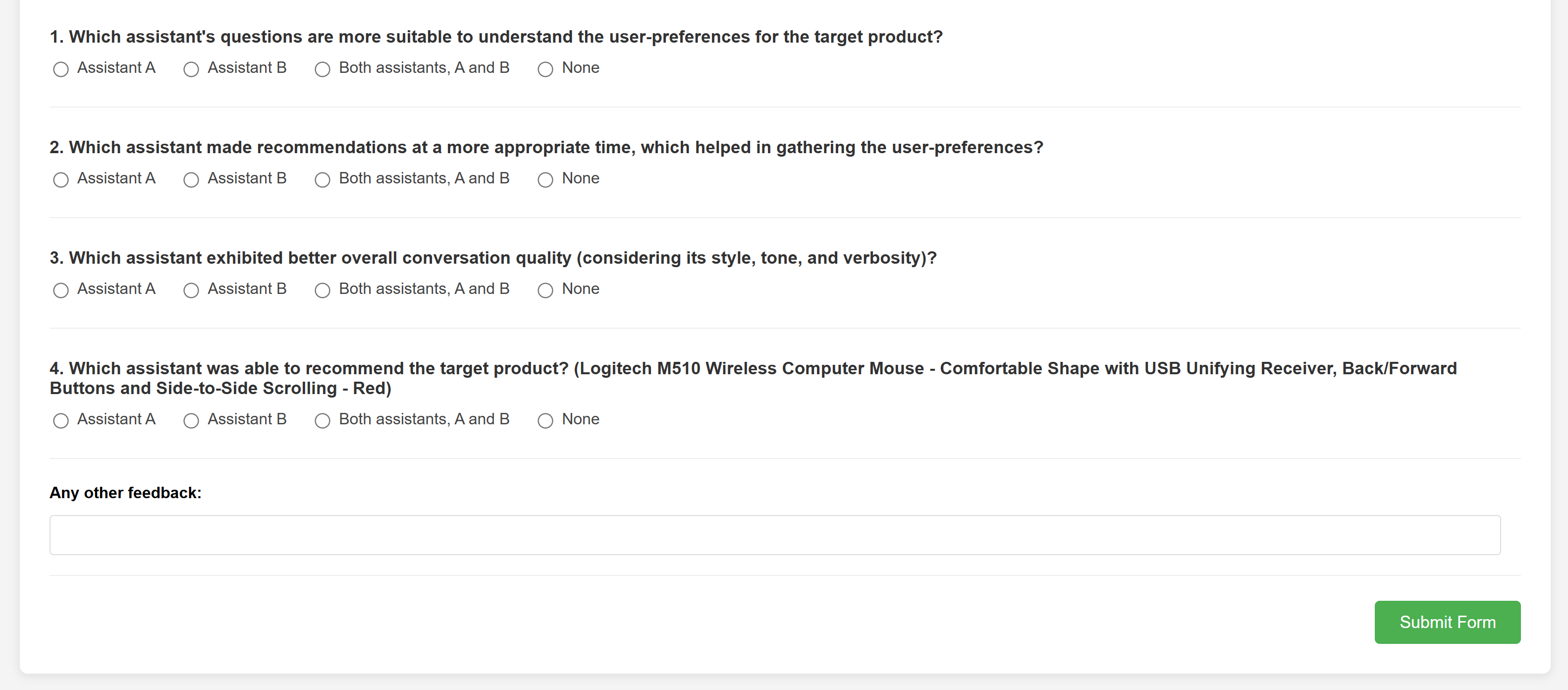}
    \caption{The participants are asked to answer 4 questions based on the displayed conversations. Additionally, the participants can provide any additional feedback they want.}
    \label{fig:userstudy-questions}
\end{figure*}

Fig. \label{sec:userStudyInst}
\ref{fig:userstudy-description} and Fig. \ref{fig:userstudy-questions} show parts of the form as visible to the participants of the user study.

\subsection{User Study: Preference between Assistant Responses}
The Table~\ref{tab:compact-userstudy} shows the preferences of the participants between the responses of different assistants.

\begin{table}[t]
\centering
\small
\setlength{\tabcolsep}{4pt}
\renewcommand{\arraystretch}{0.7}
\caption{
User responses grouped by D2D and LLM success (\cmark), failure(\xmark), \textbf{weighted} by stratification ratios. Each block shows preferences for Q1–Q3: D2D, LLM, Both, or None.
}
\label{tab:compact-userstudy}
\resizebox{0.8\linewidth}{!}{
\begin{tabular}{cc|c|cccc}
\toprule
\textbf{D2D} & \textbf{LLM} & \textbf{Ques} & \textbf{D2D} & \textbf{LLM} & \textbf{Both} & \textbf{None} \\
\midrule

\multirow{3}{*}{\cmark} & \multirow{3}{*}{\cmark}
& Q1 & 6 & 3 & 1 & 0 \\
& & Q2 & 7 & 2 & 1 & 0 \\
& & Q3 & 8 & 2 & 0 & 0 \\

\midrule

\multirow{3}{*}{\cmark} & \multirow{3}{*}{\xmark}
& Q1 & 10 & 0 & 0 & 0 \\
& & Q2 & 10 & 0 & 0 & 0 \\
& & Q3 & 8 & 0 & 2 & 0 \\

\midrule

\multirow{3}{*}{\xmark} & \multirow{3}{*}{\cmark}
& Q1 & 3 & 7 & 0 & 0 \\
& & Q2 & 3 & 7 & 0 & 0 \\
& & Q3 & 3 & 2 & 4 & 1 \\

\midrule

\multirow{3}{*}{\xmark} & \multirow{3}{*}{\xmark}
& Q1 & 13 & 7 & 0 & 0 \\
& & Q2 & 13 & 6 & 0 & 1 \\
& & Q3 & 12 & 7 & 1 & 0 \\

\midrule

\multicolumn{2}{c|}{}
& Q1 & 32 & 17 & 1 & 0 \\
\multicolumn{2}{c|}{\textbf{Total}}
& Q2 & 33 & 15 & 1 & 1 \\
\multicolumn{2}{c|}{}
& Q3 & 31 & 11 & 7 & 1 \\
\bottomrule
\end{tabular}
}
\end{table}

\end{document}